\documentclass[
12pt,
pre,
a4paper,
amsmath,
superscriptaddress,
floatfix,
notitlepage,
]{revtex4-1}

\usepackage{graphicx,amssymb,amsmath,amsfonts}
\usepackage{epsfig}
\usepackage{xcolor}
\usepackage[mathscr]{eucal}
\usepackage{comment}
\usepackage{bm}

\let \ShowFixme = 0
\let \Date = 1
\def\papertitle{Probing interface localization-delocalization transitions by colloids}

\newlength{\figwidth}

\newif\ifOnecolumn
  \ifx\Onecolumn\undefined
  \Onecolumntrue
  
\else
  \Onecolumnfalse
\fi

\newif\ifTwocolumn
  \ifOnecolumn
   \Twocolumnfalse
\else
  \Twocolumntrue
\fi

\usepackage{xspace}
\newcommand{\latin}[1]{{\it #1}}
\newcommand{\ie}{\latin{i.e.}\@\xspace}
\newcommand{\eg}{\latin{e.g.}\@\xspace}
\newcommand{\cf}{\latin{cf.}\@\xspace}
\newcommand{\etc}{\latin{etc.}\@\xspace}

\def\la{\langle}
\def\ra{\rangle}

\newcommand{\ftitle}[1]{\textbf{#1}}
\newcommand{\fig}[1]{Fig.~\ref{#1}}
\newcommand{\figs}[2]{Figs.~\ref{#1} and \ref{#2}}

\newcommand{\Fig}[1]{Figure~\ref{#1}}

\newcommand{\sect}[1]{Sect.~\ref{#1}}

\newcommand{\eq}[1]{eqn~(\ref{#1})}

\renewcommand{\H}{\mathcal{H}}
\newcommand{\m}{m}
\newcommand{\OP}{\Delta \Gamma}

\begin{document}

\title{\papertitle}

\author{Svyatoslav Kondrat}
\affiliation{Institute of Physical Chemistry, Polish Academy of Sciences, Kasprzaka 44/52, 01-224 Warsaw, Poland}

\author{Oleg A.~Vasilyev}
\author{S. Dietrich}
\affiliation{Max-Planck-Institut f{\"u}r Intelligente Systeme, Heisenbergstra{\ss}e~3, D-70569 Stuttgart, Germany}
\affiliation{IV. Institut f\"ur Theoretische Physik, Universit{\"a}t Stuttgart,  Pfaffenwaldring 57, D-70569 Stuttgart, Germany}

\begin{abstract}

	Interface localization-delocalization transitions (ILDT) occur in  two-phase fluids confined in a slit with competing preferences of the walls for the two fluid phases. At low temperatures the interface between the two phases is localized at one of the walls. Upon increasing temperature it unbinds. Although intensively studied theoretically and computationally, such transitions have not yet been observed experimentally due to severe challenges in resolving fine details of the fluid structure. Here, using mean field theory  and Monte Carlo simulations of the Ising model, we propose to detect these ILDT by using colloids. We show that the finite-size and fluctuation induced force acting on a colloid confined in such a system experiences a vivid change if, upon lowering the temperature, the interface localizes at one of the walls. This change can serve as a more easily accessible experimental indicator of the transition. 

\end{abstract}

\maketitle

\section{Introduction}

Confinement plays a key role in many physical processes and can lead to phenomena which do not occur in bulk, such as unusual drying~\cite{singh:nature:06:ConfWater} or freezing~\cite{koga:nature:01:ConfWaterFreezing, han:nphys:10:ConfWater} of confined water, an anomalous capacitance increase in subnanometer pores~\cite{gogotsi:sci:06, kondrat:jpcm:11}, a long-ranged repulsion between confined dimers~\cite{tkachenko:prl:17:LongRangeRepulsion}, \etc A particularly intensively studied system is a simple fluid, or a binary liquid mixture, confined in a slit~\cite{Fisher-et:1981, Nakanishi-et:1983, Parry-et:1990, Parry-et:1992, Binder-et:1995a, Binder-et:1995b, Binder-et:2003, schulz_binder:pre:05:LocDeloc, mueller_binder:pre:01:DelocPolymers, mueller:12:prl, milchev_et:03:prl, milchev_et:04:prl:erratum, Nowakowski-et:2009, Nowakowski-et:2008, nowak_napir:jcp:14, nowak_napir:jcp:16}. If the slit walls are identical, the critical point of the fluid is shifted, such that it approaches the bulk critical point as the slit width increases \cite{Fisher-et:1981, Nakanishi-et:1983}. However, a different \emph{type} of transition emerges if the confining walls are competing in the sense that they prefer  different phases of a simple fluid, or different species of a binary mixture~\cite{Parry-et:1990, Parry-et:1992, Binder-et:1995a, Binder-et:1995b, Binder-et:2003, schulz_binder:pre:05:LocDeloc, mueller_binder:pre:01:DelocPolymers, mueller:12:prl, milchev_et:03:prl, milchev_et:04:prl:erratum}. In this case, the transition is related to the wetting properties at a single wall, rather than to the bulk criticality of the fluid, and its location approaches the one of the wetting transition if the slit becomes macroscopically wide. A distinct feature of this transition is the emergence of an interface between the two fluid phases, which in the `ordered' phase is localized at one of the walls; in the `disordered' phase  the interface is absent and the concentration profile varies smoothly and is antisymmetric across the slit (\cf \fig{fig:deloc}b). For this reason such transitions have been called interface localization-delocalization transitions (ILDT)~\cite{Parry-et:1990, Parry-et:1992}.

Since the ILDT were predicted in 1990 by \citeauthor{Parry-et:1990}~\cite{Parry-et:1990, Parry-et:1992}, they have been intensively studied in various contexts. For instance, \citeauthor{schulz_binder:pre:05:LocDeloc}~\cite{schulz_binder:pre:05:LocDeloc} have demonstrated by Monte Carlo simulations of an Ising magnet that such transitions can be first order, \citeauthor{mueller:12:prl}~\cite{mueller:12:prl} studied them for block copolymers, and \citeauthor{milchev_et:03:prl} \cite{milchev_et:03:prl, milchev_et:04:prl:erratum} have revealed a new universality class for similar transitions in a double wedge. However, despite of the growing interest of the research community in this subject, the ILDT have not been detected experimentally yet. The closest experimental study involves X-ray scattering under grazing incidence for wetting films near a critical end point of the bulk fluid~\cite{Fukuto-et:2005}. Nuclear reaction analysis (NRA) is another experimental technique, which allows one to probe concentration profiles in films, in particular close to air. NRA was carried out in order to study the fluctuation induced increase of the width of the delocalized interface as a function of the film thickness \cite{kerle:prl:96}. However, a localization-delocalization transition has not been reported.  The reason is the challenge to resolve fine details of the spatial structure of confined fluids. Moreover, a macroscopically large system may consist of a collection of interfaces \cite{privman_fisher:83:jsp,Labbe-Laurent:17:jcp}, localized randomly at both walls with domain walls separating them. Thus, without precautious measures, experimentally it is likely to observe an average profile, resembling the phase with a delocalized interface.

Here, we study the behaviour of a \emph{colloid} suspended in a two-phase fluid and confined between competing planar and parallel walls. We shall show that within this setup the effective force acting on the colloid is sensitive to the occurrence of a phase transition inside the film. We propose that this sensitivity can be used to detect the ILDT experimentally. In order to show this, we first provide a mean-field analysis and construct the global phase diagram of a fluid in such a slit confinement (without a colloid), revealing the location of the corresponding first- and second-order transitions (\sect{sec:deloc}). Via mean-field calculations we study the effective force acting on a colloid in the vicinity of a first-order ILDT (\sect{sec:col:mft}), and confirm our mean-field results by more robust Monte Carlo simulations of the corresponding Ising model (\sect{sec:col:mft}). We summarize and conclude in \sect{sec:concl}.

\section{Interface localization-delocalization transitions}
\label{sec:deloc}

We describe a two-phase fluid by using the standard dimensionless Ginzburg-Landau-Wilson (LGW) Hamiltonian in spatial dimension $d$:
\begin{align}
\label{eq:mft:H}
	\H[\varphi] = \int_V d^d r \left\{\frac{1}{2}(\nabla \varphi)^2 + \frac{\tau}{2} \varphi^2 + \frac{g}{4!}\varphi^4 \right\} 
	+ \sum_\alpha \int_{\S_\alpha} d^{d-1} r \left\{ h_1^{(\alpha)} \varphi + \frac{c_1^{(\alpha)}}{2} \varphi^2\right\},
\end{align}
where $d^d r=\prod_{i=1}^{d} dx_\alpha$ is the $d$-dimensional volume element, $V$ is the volume accessible to the fluid, and the coupling constant $g>0$ stabilizes $\H$ below the (upper) critical point $T_c$; $\varphi(\vr)$ is the order parameter, which for an Ising magnet is the magnetization and for a binary liquid mixture the difference between the concentrations of the two species. Thus, interchangeably we shall call $\varphi$  concentration or magnetization, rather than order parameter, in order to distinguish it from the order parameter introduced below for the ILDT. Within mean field theory (MFT) one has $\tau = t/(\xi_0^+)^2$ where $t=(T-T_c)/T_c$ is the reduced temperature, and $\xi_0^+$ is the nonuniversal amplitude of the bulk correlation length $\xi=\xi_0^{+}t^{-\nu}$, $\nu=1/2$ within MFT, in the disordered phase. We focus on the ordered phase ($\tau < 0$), which corresponds to a two-phase fluid.

The second term in \eq{eq:mft:H} describes the surface contribution. Here, the sum is over the surfaces $S_\alpha$, $\alpha=\{0,w\}$, forming the slit; in \sect{sec:col}, in addition to the walls,  there will be a colloid, in which case $\alpha=\{0, w, \mathrm{col}\}$.  The surface field $h_1^{(\alpha)}$ is conjugate to $\varphi$ and $c_1^{(\alpha)}$ is the surface enhancement \cite{Diehl:1997}. Here, we focus on the fully antisymmetric case $h_1 = h_1^{(0)} = - h_1^{(w)}$ and $c_1 = c_1^{(0)} = c_1^{(w)}$. 

\begin{figure}[t]
\begin{center}
\includegraphics[width=\textwidth]{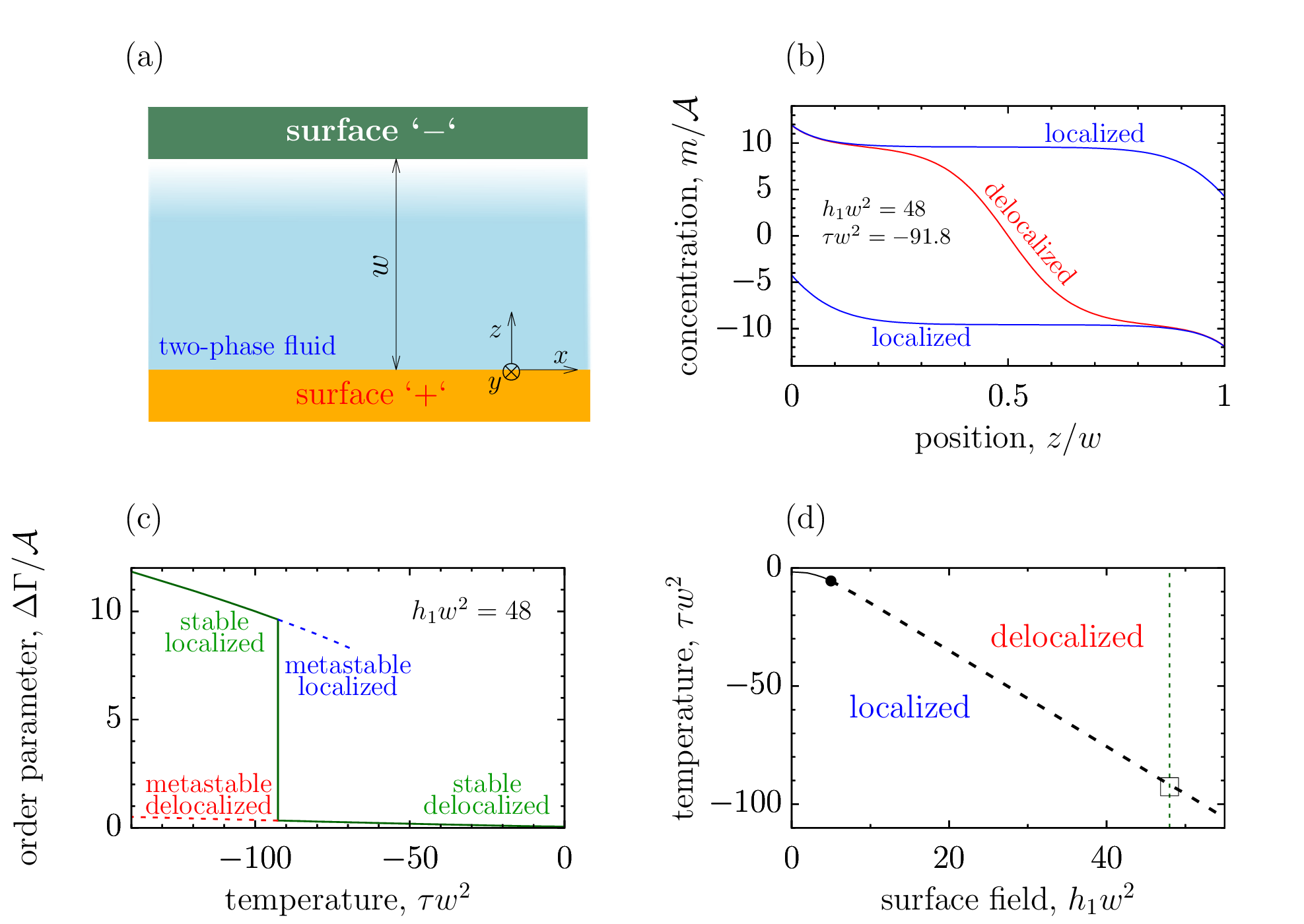}
	\caption{ \ftitle{Interface localization-delocalization transition from mean-field calculations.} (a) Schematic drawing of a model showing a two-phase fluid confined to a slit of width $w$ with competing walls. Here, the interface is located at the top wall (crossover of the shading from dark to light blue). (b) Examples of concentration profiles showing an interface located at each wall as well as the delocalized interface. The profiles are expressed in terms of the scaling function introduced in \eq{eq:m} and normalized by its amplitude $\mathcal{A} = \sqrt{6/g(\xi_0^+)^2}$ in the bulk system. The values of the surface field $h_{1} w^2=48$ and of the temperature $\tau w^2=t (w/\xi_0^+)^2= -91.8$ correspond to coexistence between localized and delocalized states (open square in (d)). In the absence of a bulk field, due to symmetry the two localized states are degenerate. (c) Order parameter $\Delta \Gamma/\mathcal{A}$ given by \eq{eq:OP} as a function of temperature for $h_{1} w^2 = 48$. Metastable branches corresponding to the localized and the delocalized interface are shown by dashed lines. (d) Phase diagram of the ILDT. The solid and the dashed lines denote second- and first- order transitions, respectively, and the filled circle shows the tricritical point. The thin dotted line shows the value of the surface field $h_1 w^2 = 48$ used in (c) and the open square denotes the value of $h_1w^2=48$ and temperature $\tau w^2 = -91.8$ used in (b). For all plots the surface enhancement is $c_1 w=1=c_1 w^{\phi/\nu} $ within MFT.
\label{fig:deloc}
}
\end{center}
\end{figure}

Minimizing \eq{eq:mft:H} with respect to $\varphi$ yields the equilibrium profile, which we express in scaling form $\m(\vr/w)$ as~\cite{Mohry-et:2010} 
\begin{align}
	\label{eq:m}
	\m (\vr/w) = \sqrt{g/6} w^{\beta/\nu} \varphi (\vr),
\end{align}
where $w$ is the slit width, and $\beta$ and $\nu$ are bulk critical exponents with the mean field values $\beta=\nu=1/2$. We shall also use the scaling variables~\cite{Mohry-et:2010} $\tau w^{1/\nu}$ for temperature, $c_1 w^{\phi/\nu}$ for the surface enhancement, and $h_1 \sqrt{g/6}w^{\Delta_1/\nu}$ for the surface field, where $\phi$ is the crossover exponent (mean field value $\phi=1/2$) and $\Delta_1$ is the surface counterpart of the bulk gap exponent with the mean-field value $\Delta_1 = 1$.

In the absence of a colloid (\fig{fig:deloc}a), the minimization problem reduces to a one-dimensional one, which in general is nevertheless difficult to solve analytically (see, however, Refs. \cite{Dantchev_2015, Dantchev_2016, Dantchev2018} for a few analytically solvable cases). Therefore, we have resorted to  numerical solutions based on our own computer code using the GNU Scientifc Library~\cite{gsl}. 

Typical concentration profiles are shown in \fig{fig:deloc}b. The profile corresponding to the delocalized interface is antisymmetric and crosses the  line $m=0$ in the middle of the slit. Besides this configuration, there are two configurations in which an interface is located at each slit wall; note that these are equally stable, degenerate states. In \fig{fig:deloc}b we show the configurations with localized and delocalized interface \emph{at coexistence}, which means that all \emph{three} configurations are thermodynamically stable (see below).

In order to characterize the ILDT, it is useful to introduce the order parameter \cite{Parry-et:1990, Parry-et:1992}
\begin{align}
	\label{eq:OP}
	\OP = \frac{1}{V} \int_V d^d r \; m (\vr),
\end{align}
which describes the excess concentration of one of the species in the slit. For a slit as in \fig{fig:deloc}, the integral in \eq{eq:OP} turns into a one-dimensional one along the $z$-axis. Since for a configuration with no interface, \ie, for the delocalized state, the concentration profile is antisymmetric (\fig{fig:deloc}b), the order parameter is zero, $\OP = 0$, while it is non-zero if there is an interface located near one of the walls.

An example of the order parameter $\OP$ as a function of temperature is shown in \fig{fig:deloc}c. It illustrates a \emph{first-order} phase transition between the two states, and the metastable branches suggest the occurrence of hysteresis. Such first-order transitions have been found previously from Monte Carlo simulations for a suitable Ising model \cite{Binder-et:1995b, Binder:pre:1996, schulz_binder:pre:05:LocDeloc}.

\Fig{fig:deloc}d shows the global phase diagram drawn in the plane spanned by the temperature $\tau$ and by the surface field $h_1$. The phase diagram consists of the lines of a second-order and a first-order phase transition, which join at a tricritical point. Our primary interest is to detect the first-order transitions; we discuss this in the next section.

\section{Detection of localization-delocalization transitions by colloids}
\label{sec:col}

We now place a colloid into the slit and study how the effective force acting on the colloid changes depending on whether the interface is located at a slit wall or delocalized. First, we carry out mean field calculations, and then we check via Monte Carlo simulations to which extent our mean field predictions conform with the results from the corresponding three dimensional Ising model.

\subsection{Mean field predictions}
\label{sec:col:mft}

Unlike for the slit walls, we consider the colloid surface to belong to the so-called extraordinary (or normal) surface universality class, which in \eq{eq:mft:H} corresponds to the limit $h_1^\mathrm{col} \to \infty$.  This implies that the magnetization diverges upon approaching the colloidal surface (see, \eg, Refs.~\cite{Binder:1983, Diehl:1986, Mohry-et:2010}). In order to deal with this numerical challenge, we have used a short-distance expansion (see Refs.~\cite{kondrat:jcp:07, Kondrat-et:2009}) for calculating the value of the magnetization at small distances $\delta$ from the colloidal surface (we took $\delta = 0.05 R$, where $R$ is the colloid radius); in the course of minimization we kept this value fixed on all mesh nodes belonging to the surface. This approach has proven to be successful in a number of previous studies~\cite{Kondrat-et:2009, Troendle-et:2009, Troendle-et:2010, mohry:softmat:14, LabbeLaurent-et:2014, Labbe-Laurent:17:jcp, vasilyev:18:sm}. Here, we have used the three-dimensional finite element library F3DM~\cite{f3dm} in order to numerically minimize the LGW Hamiltonian (\eq{eq:mft:H}) for this system.

\begin{figure}[t]
\begin{center}
\includegraphics[width=\textwidth]{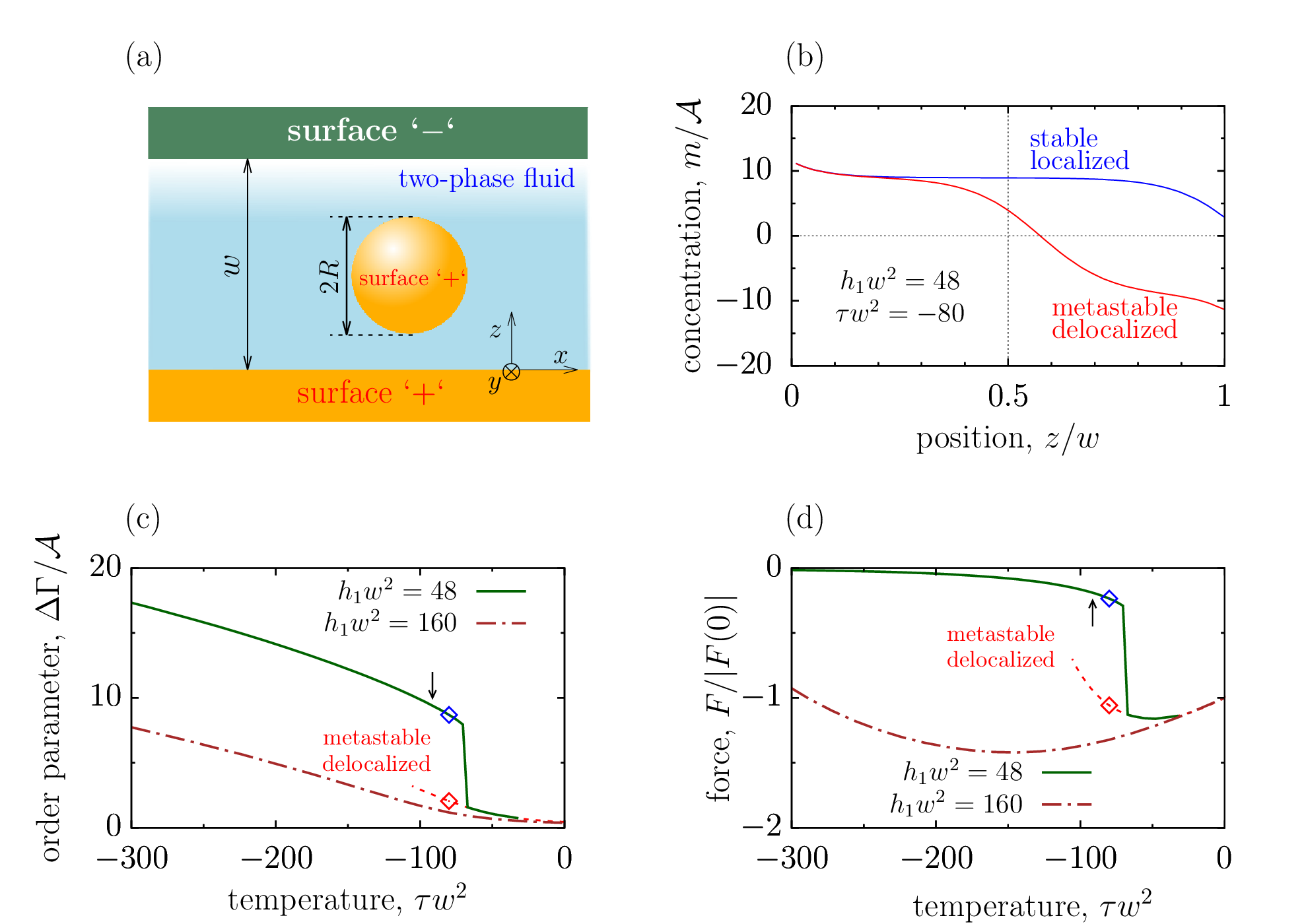}
\caption{ \ftitle{Detecting interface localization-delocalization transitions by colloids.} (a)~Schematic drawing of a model showing a colloid with radius $R$ immersed in a two-phase fluid which is confined to a slit with competing walls. (b) Examples of the concentration profiles along the $z$ axis calculated at a lateral distance $8R$ from the colloid center. The profiles are expressed in terms of the scaling function in \eq{eq:m} normalized by their amplitude $\mathcal{A}$ in the bulk system. For the values $h_1 w^2=48$ of the surface field  and $\tau w^2=t(w/\xi_0^+)^2 = -80$ of the temperature  used for this plot, the configurations with a `delocalized' interface and with the interface located at the top wall ($z=w$) are metastable and stable, respectively (see the diamonds in (c-d)). (c) Order parameter $\Delta \Gamma/\mathcal{A}$ given by \eq{eq:OP} as a function of temperature for two values of the surface field. For $h_1 w^2 = 48$ the metastable branch is shown by a dashed line, and the diamonds denote the value of $\tau w^2$ used in (b). (d) Force acting on a colloid, fixed in the middle of the slit,  as a function of temperature for two values of the surface field. The dashed line denotes the metastable branch emanating from the lower, delocalized branch of the green curve. The radius of the colloid is $R=w/4$  and the surface enhancement is taken to be $c_1 w = 1$ for all plots.  The arrows in (c) and (d) indicate the temperature of the ILDT in the absence of the colloid.
\label{fig:deloc:col}
}
\end{center}
\end{figure}

\Fig{fig:deloc:col}b shows the concentration profiles at a distance $8R$ from the colloid center. It illustrates the occurence of the localized and the delocalized interfaces also in the presence of a colloid. However, in the present case, the interface is always located at the wall with the boundary condition opposite to that of the colloid, \ie,  in Figs.~\ref{fig:deloc:col}a-b  the interface is at the top wall ($z=w$). For the range of parameters studied, we have not observed configurations with the interface at the bottom wall. This behaviour is expected because the presence of a colloid renders this state either strongly metastable or unstable. We note that the profile corresponding to a delocalized interface does not cross the  line $\m=0$ in the middle of the slit (\ie, at $z/w=0.5$), as in the case without a colloid (\fig{fig:deloc}b). Instead, it is shifted towards the wall with the preference being opposite to that of the colloid; clearly this crossing point moves towards $z=w/2$ as the distance from the colloid increases (not shown).

Interestingly, the colloid shifts the location of the first-order ILDT towards the critical point $\tau=0$ (see  \fig{fig:deloc:col}c and the arrow therein which indicates the transition temperature in the absence of the colloid). This is the case because the colloid strongly stabilizes the configurations with an interface, so that one observes not even  the \emph{metastable} states associated with them (compare Figs.~\ref{fig:deloc}c and \ref{fig:deloc:col}c; the green curve in \fig{fig:deloc:col}c does not feature a metastable branch in contrast to the green curve in \fig{fig:deloc}c). We note that due to the presence of the colloid the order parameter $\OP$ (\eq{eq:OP}) is non-zero even in the delocalized phase.

For comparison, in \fig{fig:deloc:col}c we show the order parameter $\OP(\tau)$ (see \eq{eq:OP} and the brown line) also for a surface field ($h_1 w^2=160$) for which there is no phase transition within the range of temperatures studied. In the following we shall use strong and weak surface fields in order to demonstrate the effect of the interface localization onto the force acting on such a confined colloid. In order to determine this force, we have used the stress tensor method~\cite{Kondrat-et:2009}.  In the present system, to this end, the colloid was enclosed by an ellipsoidal surface and the force was computed by integrating the stress tensor over this surface.

From  \fig{fig:deloc:col}a it is evident that the effective force acting on the colloid is negative, \ie, pointing into the negative $z$-direction, for all temperatures (\fig{fig:deloc:col}d). The `$+$` colloid surface is attracted to the `$+$` bottom surface of the slit and is repelled by the `$-$` upper slit wall so that these two force contributions add. If the fluid-fluid interface is delocalized (\ie, at high temperatures) the bottom half of the colloid is surrounded by the fluid preferred by the `$+$` surface of the slit wall, whereas the upper half of the colloid is engulfed by the phase rejected by the `$+$` field characterizing  the colloid. Together, this render a strongly negative total force. In the localized interface configuration, the upper part of the colloid is surrounded more by the fluid preferred by the colloid than in the delocalized configuration. Therefore in the latter case the repulsion from the upper slit wall is weaker than in the delocalized configuration. Accordingly, upon lowering the temperature, the total force acting on the colloid becomes less negative.

\Fig{fig:deloc:col}d shows that the ILDT manifests itself clearly in the behaviour of the force, the absolute value of which decreases sharply as the interface becomes localized at the wall.  In contrast, in the case that there is no ILDT, the force remains strong in the whole range of temperatures (brown curve in \fig{fig:deloc:col}d). This difference in the behaviour of the force can be used to detect the ILDT experimentally.

\subsection{Monte Carlo simulations}
\label{sec:col:mft}

\begin{figure}[t]
\begin{center}
\includegraphics[width=\textwidth]{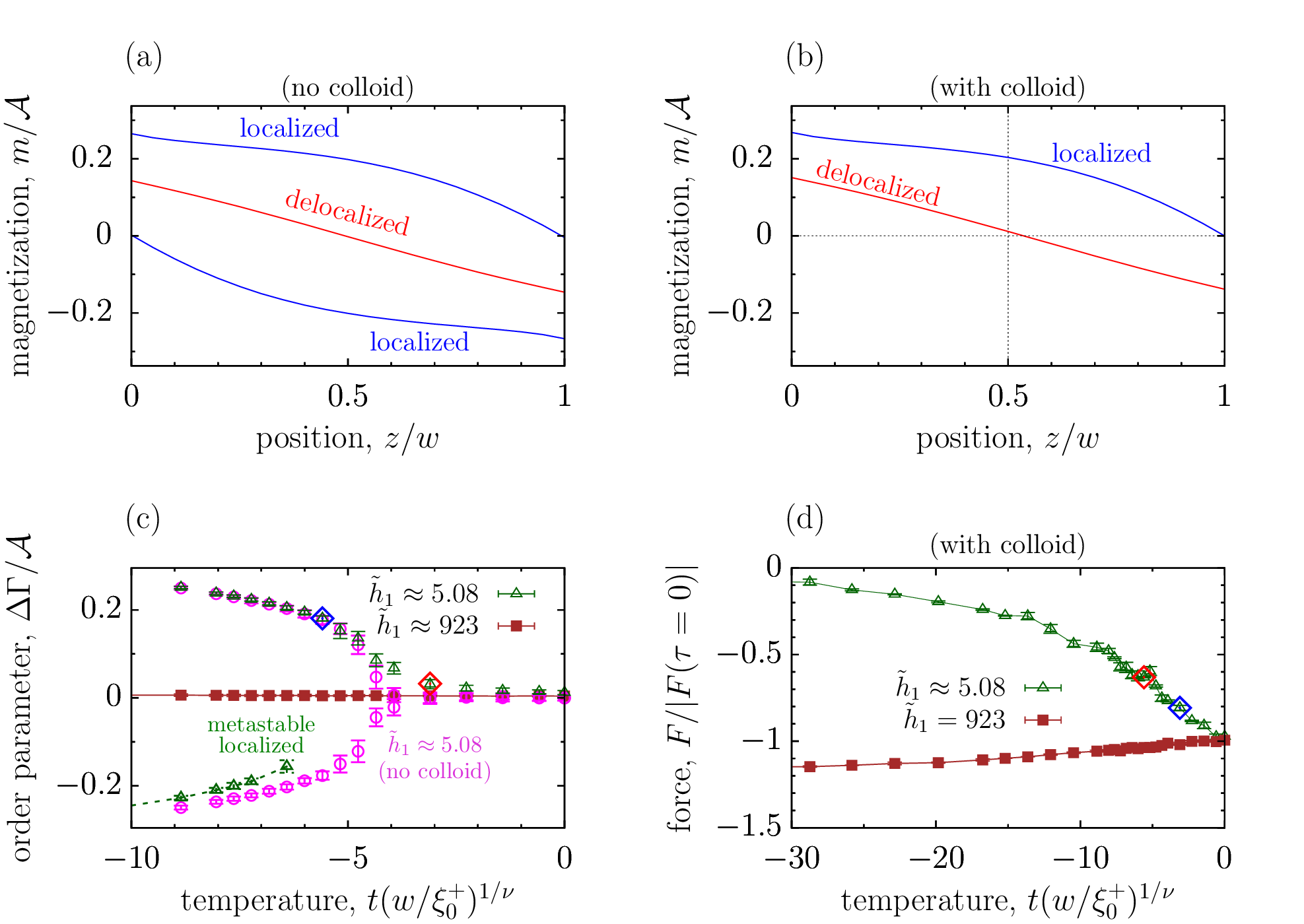}
	\caption{ \ftitle{Interface localization-delocalization transition from Monte Carlo simulations.} (a)~Examples of the magnetization profiles along the $z$ axis (across the slit) for an Ising slit with competing walls without a colloid. (b) Examples of magnetization profiles along the $z$ axis calculated at a lateral distance $11.6 R$ from the colloid center. In both (a) and (b) the surface field scaling variable is $\tilde h_1 = h_1 w^{\Delta_1/\nu} \approx 5.08$ and the temperature scaling variable is $t (w/\xi_0^+)^{1/\nu} \approx -3.0$  (delocalized) and $t (w/\xi_0^+)^{1/\nu} \approx -5.7$ (localized interfaces; the two localized configurations are degenerate). Here $t=(T-T_c)/T_c$ is the reduced temperature, $\xi_0^+$ is the correlation length critical amplitude in the high temperature phase; the scaling variable $t(w/\xi_0^+)^{1/\nu}$ for the Ising model corresponds to $\tau w^2$ for MFT (\figs{fig:deloc}{fig:deloc:col}). For these parameters, we have not found a metastable state with a throughout negative magnetization. The profiles in (a-b) are normalized by their amplitude $\mathcal{A} \approx 1.6019045$ in the corresponding bulk system \cite{talapov_bloete:jpa:96:3DIsing}. (c) Order parameter $\Delta \Gamma/\mathcal{A}$ given by \eq{eq:OP} as a function of temperature for two values of the surface field. The metastable branch for $\tilde h_1 \approx 5.08$ is shown by a dashed line. Here and in (d) the diamonds denote the values of $t (w/\xi_0^+)^{1/\nu}$ used in (a) and (b). (d) Force acting on a colloid as a function of temperature for two values of the surface field. In (b-d), the colloid of radius $R/w = 5.5/21 \approx 0.26$ is located in the middle of the slit. 
\label{fig:deloc:mc}
}
\end{center}
\end{figure}

We have also performed Monte Carlo simulations of the Ising model defined on a simple cubic lattice with lattice constant $a$, and with a classical spin $\sigma_{i}= \pm 1$ assigned to each lattice site $i$, which models an incompressible binary liquid mixture or a simple fluid. We consider a system of $M=N_{x} \times N_{y} \times N_z=344064$ ($N_x=128$, $N_y=128$, $N_z=21$) lattice sites, which is periodic in the $x$ and $y$ directions, but has a width $w=N_z a$ in $z$ direction. The Hamiltonian of such an Ising slit is given by
\begin{align}
\label{eq:ham_b}
H(\{ \sigma \})= -J \sum \limits_{\la ij \ra }\sigma_{i} \sigma_{j}
	+ \sum_{\alpha} h_{1}^{\alpha} \sum \limits_{ k \in S_\alpha }  \sigma_{k},
\end{align}
where the coupling constant is $J=1$ if the temperature $T$ is measured in units of $J/k_B $. The sum $\la ij \ra$ runs over all neighboring pairs of spins, and the second sum is over the spins at the slit walls $S_{\alpha,w}$ (and the colloidal surface $S_{col}$). The colloid (if any) is defined by drawing a sphere of radius $R$ with its center located at a lattice site, and all spins within this sphere are treated as being frozen; the total number of fluctuating spins reduces accordingly)~\cite{vasilyev:pre:2014:colloidsMC}. The simulations correspond to the ratio $R/w=5.5/21\simeq 0.26$ of the colloid radius to the slit width.

The Monte Carlo simulations have been performed by using $10^6$ hybrid MC steps. Each MC step consists of a combination of a Wolff cluster update and $M/2$ single spin Metropolis updates. In order to capture metastable configurations, we start from setting the spins to $+1$ (or spin $-1$) on all sites as an initial configuration; then we use $M$ Metropolis updates per MC step without Wolff updates, which prevents the system from fluctuating between different (meta)stable states. The force has been calculated by computing the difference between the insertion free energies of the colloid in two neighboring positions (in the middle of the system at $z=11a$ and at $z=10a$) using the local order parameter integration method; for details see~Refs.~\cite{vasilyev:pre:2014:colloidsMC, Vasilyev_Maciolek:2015, vasilyev:18:sm}. For the surface field at the slit walls we have taken the value $h_1 = h_1^{(0)} = - h_1^{(w)}=0.55$ (which corresponds to the surface field scaling variable $\tilde h_1 = h_1 w^{\Delta_1/\nu} \approx 5.08$, with the correlation length critical exponent $\nu \approx 0.63002(10)$~\cite{Hasenbusch-nu} and the surface field exponent $\Delta_1 = 0.46(2)$~\cite{Guida_1998}). As has been shown previously~\cite{Binder-et:1995b,Binder:pre:1996}, in the absence of the colloid this value of the surface field renders the ILDT to be first order. For the colloid we have used an infinite surface field, which amounts to fixing the spins on the lattice sites neighboring the sites of the colloid surface. In the following we use the temperature scaling variable $t (w/\xi_0^+)^{1/\nu}$, with the reduced temperature $t=(T- T_c)/T_c=(\beta_c-\beta)/\beta$ and the inverse temperature in units of $J/k_B$ of the critical point $\beta_c =1/T_c\simeq 0.22165455(3)$ \cite{Deng:03:pre}. Measured in units of $a$, the value of $\xi_0^+$ is $\xi_0^+=0.501(2)$~\cite{Ruge-et:1994}. 

The results of our simulations are provided in \fig{fig:deloc:mc}. Figure~\ref{fig:deloc:mc}a shows the magnetization profile $\left<\sigma_i \right>$ corresponding to the localized and delocalized configurations \emph{in the absence} of a colloid.  Figure~\ref{fig:deloc:mc}b demonstrates that such configurations occur also when a colloid (of radius $R/w =5.5/21 \approx 0.26$) is placed and fixed in the middle of the slit. Similarly as in the mean-field calculations, the presence of a colloid renders one of the two configurations with a localized interface unstable (or highly metastable). However, different from MFT, we \emph{have} obtained metastable configurations with an interface located at the wall with the same preference as the colloid, but we have not been able to locate a metastable branch with a delocalized interface (compare Figs.~\ref{fig:deloc:col}c and \ref{fig:deloc:mc}c). The former observation comes about by starting the MC simulation with an initial configuration in which all spins are negative. For suitable temperatures the aforementioned interface emerges in the course of the simulation near the `$+$` wall and fluctuates there, leaving the colloid in the `$-$` phase.

Similarly as predicted by MFT, the absolute value of the force acting on the colloid drops upon approaching  the ILDT from the delocalized states (\fig{fig:deloc:mc}d), although its change is smoother and appears to be not discontinuous as in MFT (Figs.~\ref{fig:deloc:mc}c-d). This is likely because of a rounding-off of the transition, which is due to the presence of the colloid and due to finite size effects. However, we compare this force with the one obtained for a very strong surface field, for which no transition occurs within the range of temperatures considered (\fig{fig:deloc:mc}c-d). In this case, the force remains to be strong as the temperature decreases, confirming the MFT prediction that a colloid can be used to detect the occurrence of  ILDT.

\section{Conclusion}
\label{sec:concl}

We have shown that colloids can be used to detect the interface localization-delocalization transition (ILDT) occurring for a two-phase fluid confined between two parallel walls, which have opposing preferences for the two phases. The ILDT is characterized by the emergence of a  localization of the fluid-fluid interface at one of the walls. While intensively studied theoretically and computationally~\cite{Parry-et:1990, Parry-et:1992, Binder-et:1995a, Binder-et:1995b, Binder-et:2003, schulz_binder:pre:05:LocDeloc, mueller_binder:pre:01:DelocPolymers, mueller:12:prl, milchev_et:03:prl, milchev_et:04:prl:erratum}, this transition has not yet been observed experimentally due to challenges in resolving the fluid structure and, in a macroscopically large system, due to jumps of the interface   between the two walls, forming domains.

The envisaged role of the colloid is two-fold. On one hand, it stabilizes the interface at one wall, and renders the localization of  the interface at the other, opposite wall unstable or highly metastable. On the other hand, the colloid experiences a vivid change in the force acting on it in the vicinity of a (first-order) ILDT. We show by mean-field calculations and Monte Carlo simulations that upon  lowering of the temperature the absolute value of the force drops  significantly at the transition (see Figs.~\ref{fig:deloc:col}d and \ref{fig:deloc:mc}d, respectively). This change in the behaviour of the force can serve as an experimental indicator of the transition.

Our Monte Carlo simulations predict the force acting on a colloid with radius $R=5.5a$ in a slit of  width $w=21a$ to be of the order of a few $k_BT$ ($\approx 4$ pN$\times$nm at room temperature) per lattice constant $a$. For the Ising model the amplitude of the correlation length is $\xi_{0}^{+} \approx 0.501a$, while it is of the order of a nanometre for a typical binary liquid mixture; for example, for a 3-methyl-pyridine (3MP)/heavy water mixture it is $\xi_0^+ \approx 1.5$nm~\cite{Dang-et:2013} and for water and 2,6-lutidine $\xi_0^+=0.2$nm~\cite{Gambassi-et:2009b}. Thus the critical force is of the order of a few pN. Under favorable conditions, recently employed experimental methods allow one to measure it also in the presence of background forces, \eg,  electrostatic ones or van der Waals forces~\cite{Gambassi-et:2009b,Gambassi-et:2009}. Such measurements can be accomplished by a number of techniques. For instance, \citeauthor{Hertlein-et:2008}~\cite{Hertlein-et:2008} have measured the force between a colloid and a substrate using total internal reflection microscopy (TIRM). Another possibility consists of holographic optical tweezers combined with digital video microscopy~\cite{paladugu:15:nonaddcas:arxive} and laser-scanning confocal microscopy~\cite{Nguyen-et:2013, stuij_maciolek_schall:sm:17:CCFColl}. Although such measurements have been carried out in the mixed (disordered) phase of a critical binary liquid mixture, it would be promising to adopt them to the measurements in the demixed (two-phase) region in order to probe the above predictions experimentally.

Finally, we emphasize that our calculations do not indicate any easily observable changes in the force acting on the colloid in the case of a \emph{second}-order ILDT. One possibility to detect this transition might be to study \emph{fluctuations} of the force, which are expected to increase in the localized state due to jumping between the two (meta)stable states. In experiments, fluctuations of the colloid position are measured to infer an interaction potential \cite{Hertlein-et:2008, paladugu:15:nonaddcas:arxive}, and this can provide information on the force fluctuations. Within our simulation framework, however, such fluctuations are not straightforward to measure. In order to calculate the force acting on a colloid, the insertion free energy of the colloid is computed by integrating the local magnetization over the local field \cite{vasilyev:pre:2014:colloidsMC}. This makes it difficult to compute the force as a function of `time', which is needed to estimate the force fluctuations. It will be interesting to explore this possibility in future research.

\begin{acknowledgements}
	
	SK was supported in part  by the European  Unions  Horizon  2020 research and innovation programme under the Marie Sk{\l}odowska-Curie grant agreement No.~734276. We thank an anonymous referee for raising the issue of force fluctuations.

\end{acknowledgements}

\bibliography{deloc}

\end{document}